\documentclass{aastex62}
\usepackage{graphicx}
\graphicspath{{figures/}}
\usepackage{amsmath}
\usepackage{amssymb}
\usepackage{color}

\submitjournal{AJ}
\accepted{August 16, 2018}

\shorttitle{Europa's Optical Aurora}
\shortauthors{de Kleer and Brown}

\begin{document}

\title{Europa's Optical Aurora}

\correspondingauthor{Katherine de Kleer}
\email{dekleer@caltech.edu}

\author[0000-0000-0000-0000]{Katherine de Kleer}
\affil{California Institute of Technology \\
1200 E California Blvd M/C 150-21 \\
Pasadena, CA 91125, USA}

\author{Michael E. Brown}
\affil{California Institute of Technology \\
1200 E California Blvd M/C 150-21 \\
Pasadena, CA 91125, USA}



\begin{abstract}
Auroral emissions provide opportunities to study the tenuous atmospheres of Solar System satellites, revealing the presence and abundance of molecular and atomic species as well as their spatial and temporal variability. Far-UV aurorae have been used for decades to study the atmospheres of the galilean satellites. Here we present the first detection of Europa's visible-wavelength atomic oxygen aurora at 6300/6364 \AA{} arising from the metastable O$(^1$D) state, observed with the Keck I and Hubble Space Telescopes while Europa was in eclipse by Jupiter on six occasions in February-April 2018. The disk-integrated O($^1$D) brightness varies from $<$500 R up to more than 2 kR between dates, a factor of 15 higher than the OI 1356 \AA{} brightness on average. The ratio of emission at 6300/5577 \AA{} is diagnostic of parent molecule; the 5577 \AA{} emission was not detected in our dataset, which favors O$_2$ as the dominant atmospheric constituent and rules out an O/O$_2$ mixing ratio above 0.35. For an O$_2$ atmosphere and typical plasma conditions at Europa's orbit, the measured surface brightness range corresponds to column densities of 1-9$\times$10$^{14}$ cm$^{-2}$. 
%
%
\end{abstract}



\section{Introduction} \label{sec:intro}
Europa's tenuous atmosphere is sourced from sputtering of its surface by energetic particles in the jovian environment and is therefore indicative of  surface composition. The surface is composed of H$_2$O ice coated with darker materials that follow a distinct spatial distribution (McCord et al. 1999). Some species may have originated in Europa's subsurface ocean and been carried up through the ice shell via upwelling; a greater understanding of the chemical make-up of these materials could shed light on the conditions within the ocean. In addition, if geological processes such as cryovolcanism are active today, the release of volatiles into the atmosphere may produce detectable atmospheric signatures. Europa's exospheric composition and variability thus provide information on surface composition, the presence and nature of active geophysical processes, and chemical processes driven by solar radiation and interactions with the local particle environment. \par
Europa's atmosphere was first detected by Hall et al. (1995) from far-UV oxygen emissions; the ratio in intensity of the 1356 and 1304 \AA{} OI emission lines was used to identify O$_2$ as the dominant atmospheric constituent, and the line strength yielded a column density of (2.4-14)$\times$10$^{14}$ cm$^{-2}$ (Hall et al. 1998). In 1999 and 2012-2015, Roth et al. (2014; 2016) observed Europa's far-UV auroral emissions over 70 Hubble Space Telescope (HST) orbits comprising 20 distinct visits. They found similar atmospheric properties: an O/O$_2$ mixing ratio of $\lesssim$1-5\% and a typical column density of 9$\times$10$^{14}$ cm$^{-2}$ when using a matching electron density of 40 cm$^{-3}$. They show that the brightness and morphology of Europa's aurorae are variable by a factor of $\sim$5, which they attribute largely to the position of Europa relative to the jovian plasma environment.\par
In their December 2012 dataset, Roth et al. (2014) detected localized off-limb hydrogen and oxygen emissions near Europa's south pole, which they used to infer an active water plume. A lack of similar detections in the other 19 visits indicates that such events are rarely detectable in this type of observation, though subsequent work by other groups using a variety of methods support the presence of sporadic plume activity originating from a few distinct surface locations (Sparks et al. 2016; 2017; McGrath and Sparks 2017; Jia et al. 2018). \par
Beyond atomic O and the single detection of H, from which the presence of O$_2$ and H$_2$O are inferred, no other species have been seen in Europa's near-surface atmosphere. However, atomic hydrogen has also been detected in an extended corona through absorption of Jupiter's light during Europa transit (Roth et al. 2017). In addition, both Na and K have been observed in Europa's extended atmosphere ($>5R_{Europa}$); the relative abundance of these species is different from that of Io and of micrometeorites, suggesting an endogenic origin of this alkali material (Brown and Hill 1996; Brown 2001). These species were observed at visible wavelengths, where the reflected sunlight off of Europa's high-albedo disk drowns out the faint gas signatures within a few Europan radii of the surface; detecting Europa's visible-wavelength auroral emissions requires observing in the absence of reflected sunlight, which in Earth-based observations requires observing when Europa is in Jupiter's shadow. \par
We observed Europa in eclipse by Jupiter on six occasions in the spring of 2018 with high-resolution optical spectroscopy from HST and Keck Observatory in order to measure Europa's auroral emissions, which have not previously been detected at these wavelengths. The observations are sensitive to wavelengths where O, H, and Na emit, and hence provide an independent characterization of Europa's oxygen atmosphere and a complementary avenue for searching for water plume activity. These two telescopes provide complementary features: HST is able to spatially resolve the emission and hence constrain morphology, while the Keck observations cover a wide wavelength range and have the potential to detect new species at Europa. In this article we present results from observations during five eclipses observed with HST and one eclipse observed with Keck, between February and April 2018. Section \ref{sec:obs} describes the observations and data reduction procedures, and the results are presented in Section \ref{sec:results}. The emission model is described in Section \ref{sec:model}, and implications of the data in the context of the modeling results are discussed in Section \ref{sec:disc}. The conclusions are summarized in Section \ref{sec:conc}.
%
%
%
\section{Observations and Data Reduction} \label{sec:obs}
Our observations target Europa's optical auroral emissions; Europa's disk is bright at optical wavelengths, and detecting emission from the tenuous atmosphere at these wavelengths requires observing while Europa is in Jupiter's shadow. We observed Europa during six eclipses by Jupiter in February-April 2018 with HST (five eclipses) and the Keck I telescope (one eclipse). The observations are described below and summarized in Table \ref{tbl:obs}. \par
\begin{table}
\caption{Observations\label{tbl:obs}}
\centering
\begin{tabular}{ccccccccc}
\hline
Date & Tel/Inst & Slit size & R & Wavelength & Jup. Limb Dist. & Europa Diam. & CML & On-source time \\
$[\mathrm{UT}]$ & & [arcsec] & & [\AA{}] & [arcsec] & [arcsec] & [$^{\circ}$W] & [sec] \\
\hline 
2018-02-18 & HST/STIS CCD & 52$\times$2 & 5000 & 5808-6867 & 8-13 & 0.83 & 353 & 568 per setting \\
2018-02-25 & HST/STIS CCD & 52$\times$2 & 5000 & 5808-6867 & 14-19 & 0.84 & 351 & 456 per setting \\
2018-03-01 & HST/STIS CCD & 52$\times$2 & 5000 & 5808-6867 & 6-12 & 0.85 & 354 & 612 per setting \\
2018-03-04 & HST/STIS CCD & 52$\times$2 & 5000 & 5808-6867 & 19-25 & 0.86 & 350 & 568 per setting \\
2018-03-22 & Keck I/HIRES & 7$\times$1.7 & 25,000 & 5400-9600 & 0-23 & 0.91 & 351 & 1200 usable\\
2018-04-05 & HST/STIS CCD & 52$\times$2 & 5000 & 5808-6867 & 10-16 & 0.94 & 353 & 516 per setting\\
\hline
\end{tabular}
\end{table}

\subsection{HST STIS}
We observed Europa during five eclipses with the Space Telescope Imaging Spectrograph (STIS) CCD on HST; each eclipse was observed over a single HST orbit. Observations were made with the 52''$\times$2'' slit and the G750M medium-resolution (R$\sim$5000) grating in two spectral settings centered on 6581 and 6094 \AA{}, which cover 6295-6867 and 5808-6380 \AA{} respectively. The former setting covers H$\alpha$ emission at 6563 \AA{}, while the latter covers OI emission at 6300/6364 \AA{} and the sodium D lines at 5889/5896 \AA{}. \par
During each eclipse, we obtained images in two dither positions, and split each integration into two to facilitate cosmic ray identification and removal. All observations were reduced, rectified, spectrally and photometrically calibrated, and corrected for hot pixels and cosmic rays using the standard \textbf{calstis} pipeline. Spatially-varying scattered light from Jupiter dominates the signal in the final pipeline data products, and we test two independent techniques to remove the scattered light: (1) Interpolation and subtraction of the scattered light background based on the average spectral and spatial profile of the background; and (2) Pair subtraction of the dithered images after scaling the median brightness to match in the spatial region surrounding Europa. The latter method resulted in a better removal of artifacts in the data, producing a more gaussian noise distribution; this method was used to produce the final calibrated and cleaned images. \par
Europa's location on the detector was determined by inspecting the pair-subtracted image for each dither pair and identifying the characteristic positive and negative signal pair. The auroral emission from Europa is faint and does not cover the whole satellite, and the center of Europa's disk could therefore not be determined from the images to better than 5 pixels (0.25''). We tested extracting Europa's brightness using apertures offset from the nominal pointing by up to 10 pixels both along and across the slit, but found that the nominal pointing maximized the signal to noise of the 6300 \AA{} emission, and we therefore assumed the nominal pointing for our analysis.\par
Europa's surface brightness in each dithered image was extracted using an aperture the size of Europa's disk on the date of observation. The surface brightness within this aperture was averaged to produce the disk-integrated brightness of each emission line on each date. The quoted uncertainty on surface brightness is the standard deviation of the brightnesses derived by applying the identical aperture-extraction procedure to regions of the spectrum free of emission features. \par
%
\subsection{Keck HIRES}
%
%
One additional eclipse was observed with the High Resolution Echelle Spectrometer (HIRES; Vogt et al. 1994) on the Keck I telescope. The observations were made with the 7$\times$1.7'' slit, which covers Europa's $\lesssim$1'' disk completely, and achieve a resolving power of R$\sim$25,000 over the 5400-9600 \AA{} wavelength range. Europa has no detectable continuum emission in eclipse at these wavelengths, and pointing and tracking were performed by offsetting from a nearby satellite and tracking at Europa's ephemeris rate, which was updated at the start each integration. After every 300-second integration of Europa, we offset to the guide satellite to check pointing and recenter the satellite in the slit if needed before offsetting back to Europa.\par
The data were corrected for bias level and flat fielded. Image rectification was performed manually for each order based on fitting a polynomial trace to the continuum spectrum of the guide satellite. Thorium-Argon lamp observations were performed for wavelength calibration, and sky lines provided an additional check on calibration over much of the spectrum. \par
Thick and variable cloud cover obscured Europa in many of the integrations. Oxygen emission from Europa at 6300 and 6364 \AA{} was detected through brief breaks in the clouds in half of the 14 integrations; of these, we select the four with the lowest Jupiter background and the best centering within the slit to produce the final spectra. The spectra are dominated by Jupiter's scattered light at the small angular separation of Europa during eclipse. Average spatial and spectral profiles for the background are computed for each order individually to construct an artificial background image that is subtracted from the data. \par
Although weather conditions were not photometric, we perform a rough calibration using observations of Jupiter's meridian taken with the same instrument and on a later night during clear conditions. We use these data to derive a conversion between detector units (counts/sec) and Jupiter's known brightness of 4.7 MR/\AA{} at 6300 \AA{} (for a Sun-Jupiter distance of 5.422 AU; Woodman et al. 1979), which is used to calibrate the Europa data. \par
The data are integrated over wavelength and over a 1.8'' interval along the slit and divided by the angular size of Europa to produce a final Europa emission spectrum in Rayleighs (1 R = 10$^6$/4$\pi$ photons/s/cm$^2$/steradian). The derived values vary by a factor of five between observations in which Europa was detected, likely due to both variable cloud cover and the intrinsic variability of the aurora. The spectrum with the highest detected emission corresponds to the period when the sky conditions were clearest, and we use this integration for our estimates of Europa's auroral emission on this date. Due to the fact that the Jupiter observations used to calibrate the data were made on a clearer night than the target observations, this estimate is a lower limit. \par
\section{Results} \label{sec:results} \label{sec:disc}
We detected OI emission from Europa at 6300/6364 \AA{} in the Keck observations and in four of the five HST observations; the date when OI emission was not detected coincides with the closest proximity of Europa to Jupiter (see Table \ref{tbl:obs}) and hence the highest background noise due to scattered Jupiter light. The disk-integrated auroral brightness on each date is given in Table \ref{tbl:brightness}; the brightness ratio of the 6300 and 6364 \AA{} lines is consistent within 1.5$\sigma$ on all dates with the ratio of three expected for an optically-thin source.\par
\begin{figure}[ht]
\centering
\includegraphics[width=14cm]{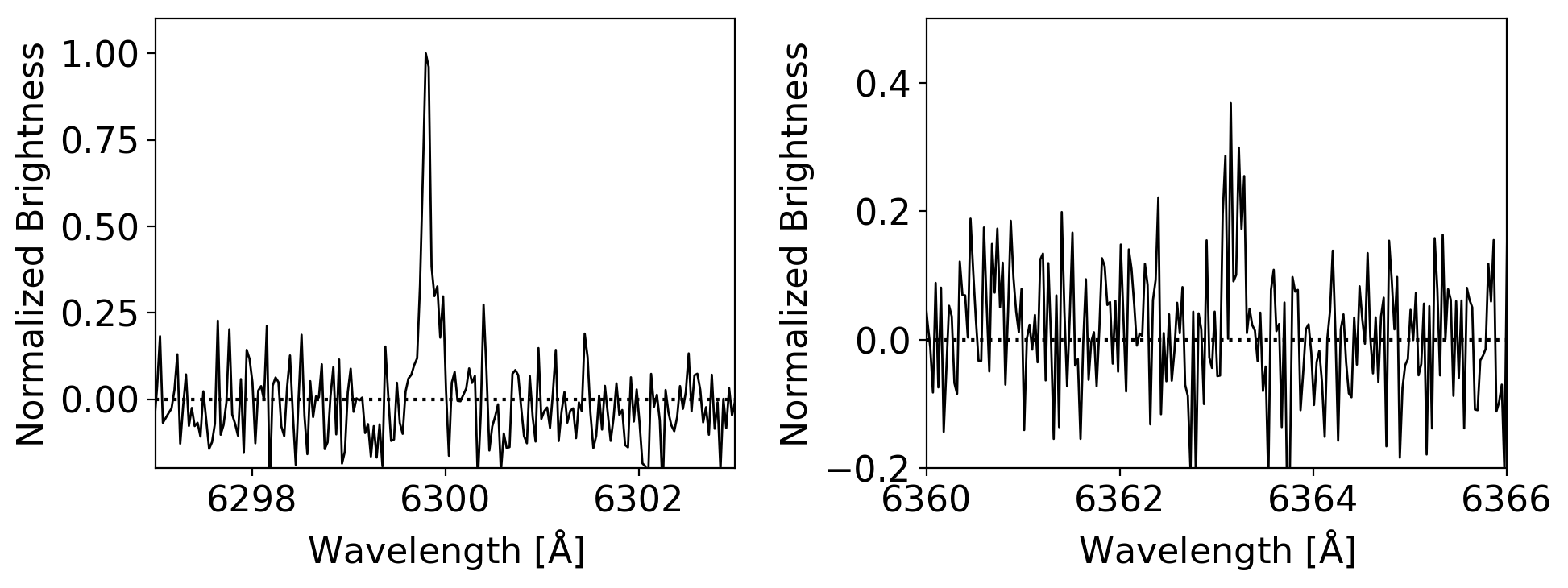}
\caption{Keck/HIRES spectrum of emission at 6300 and 6364 \AA{} derived from the frames with the least cloud obscuration. Both spectra are normalized to the peak of the 6300 \AA{} line. \label{fig:spectra}}
\end{figure}

Figure \ref{fig:spectra} shows the spectrum of 6300 and 6364 \AA{} emission derived from the Keck observations, in which Europa is not spatially resolved. Due to poor weather conditions during the observations we are only able to place a lower limit on the disk-integrated auroral brightness. From the four clearest integrations averaged together, we measure surface brightnesses of 135 R at 6300 \AA{} (SNR=11) and 48 R at 6364 \AA{} (SNR=5); the ratio of emission between the 6300 and 6364 \AA{} lines matches well with the expected ratio of 69/22 for emission from an optically-thin source. For the single integration with the clearest sky conditions, we measure brightnesses of 190 R and 45 R for the emission at 6300 and 6364 \AA{} respectively, and we use these numbers as the lower limits on the total auroral brightness on this date. The quoted SNR in each case is based on the standard deviation of the brightnesses derived exactly as the Europa brightness is derived, across the spectral order that contains that emission line. However, note that the noise level implied by these numbers does not represent the uncertainty on the absolute auroral brightness due to the uncertainty in the flux calibration. We do not detect any other species, and place 1$\sigma$ a upper limit of 10 R on both H$\alpha$ and O($^1$S) 5577 \AA{}, but note that these upper limits are subject to uncertainties in the flux calibration. \par
%
\begin{figure}[ht]
\centering
\includegraphics[width=18cm]{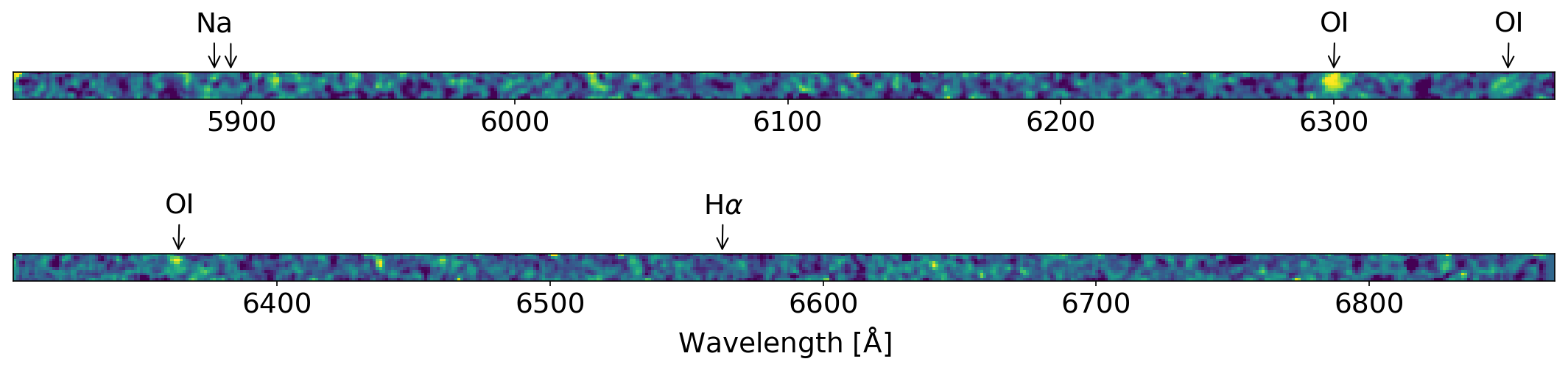}
\caption{HST/STIS spectra averaged across all five dates of observation, binned to a resolution of 0.1'' and smoothed. The two spectral settings overlap, and 6364 \AA{} OI emission is visible in both. The OI emissions at 6300 and 6364 \AA{} are indicated, as are the locations of Na and H$\alpha$ emission although these species are not detected. The auroral emissions can be distinguished from noise of comparable localized brightness by their larger spatial extent. \label{fig:fullspecHST}}
\end{figure}

Figure \ref{fig:fullspecHST} shows the full spectral images from the HST observations in the vicinity of Europa averaged across all dates, demonstrating the clear detection of 6300 and 6364 \AA{} OI emission and the non-detection of H$\alpha$ and sodium.  With all five observations combined (total integration time of 45 minutes per filter), we place a 1$\sigma$ upper limit on the disk-averaged surface brightness of any non-detected emissions of 130 R at 6295-6867 \AA{} and 170 R at 5808-6380 \AA{}. These limits are derived from the standard deviation of surface brightness averaged over the size of Europa at all wavelengths with no detected emission; the higher limit for the latter wavelength range is due to the fact that these observations were made when Europa was closer to Jupiter. \par
\begin{figure}[ht]
\centering
\includegraphics[width=8cm]{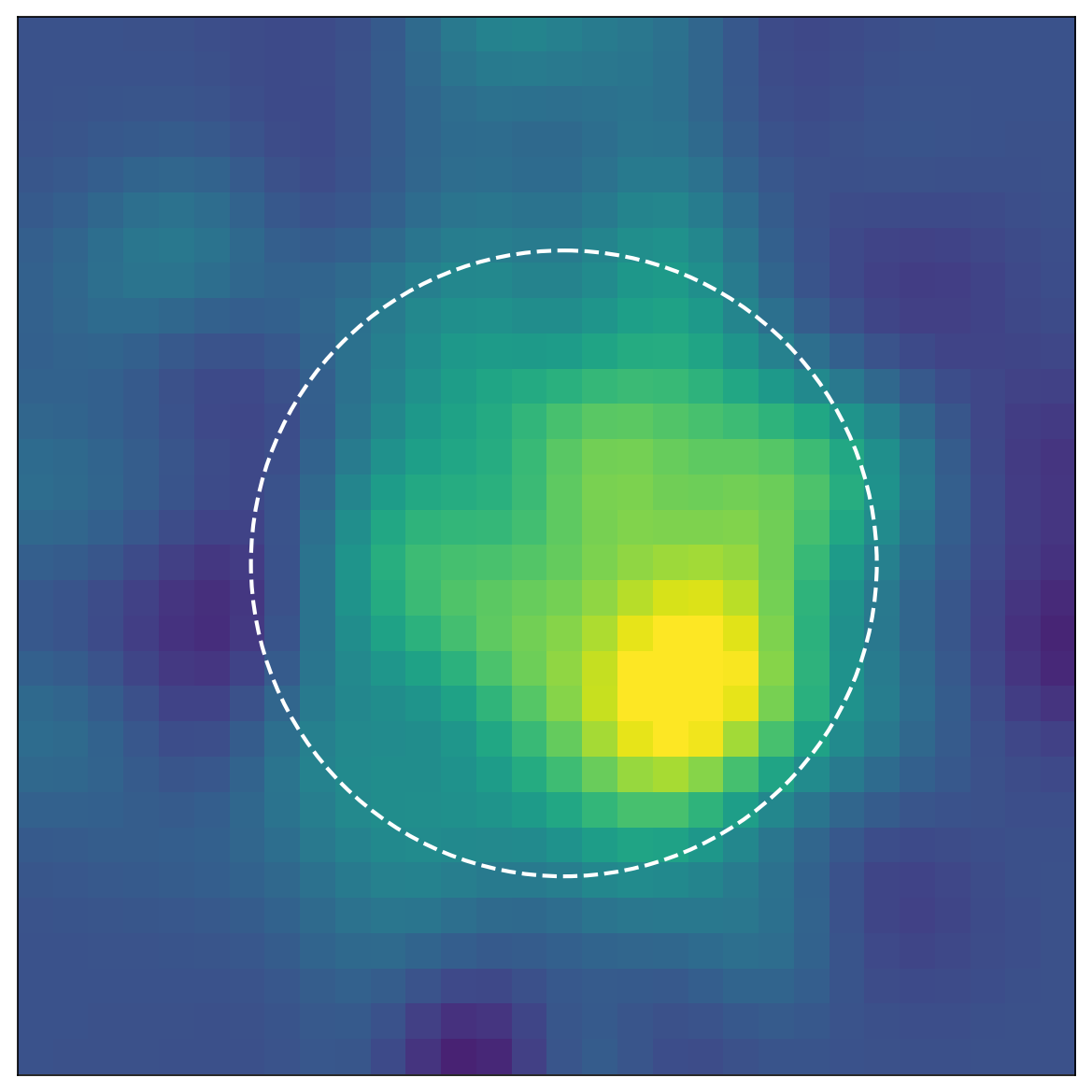}
\caption{HST/STIS image of Europa's oxygen aurora, averaged across all five visits and both 6300 and 6364 \AA{} and smoothed to bring out coherent structures. Europa's north pole is up, and more emission is seen over the trailing/dusk side. The images were not stretched to match Europa's changing radius before being combined, and the dashed circle represents the average size of Europa over the five visit dates. \label{fig:bigim}}
\end{figure}
\begin{figure}[ht]
\centering
\includegraphics[width=16cm]{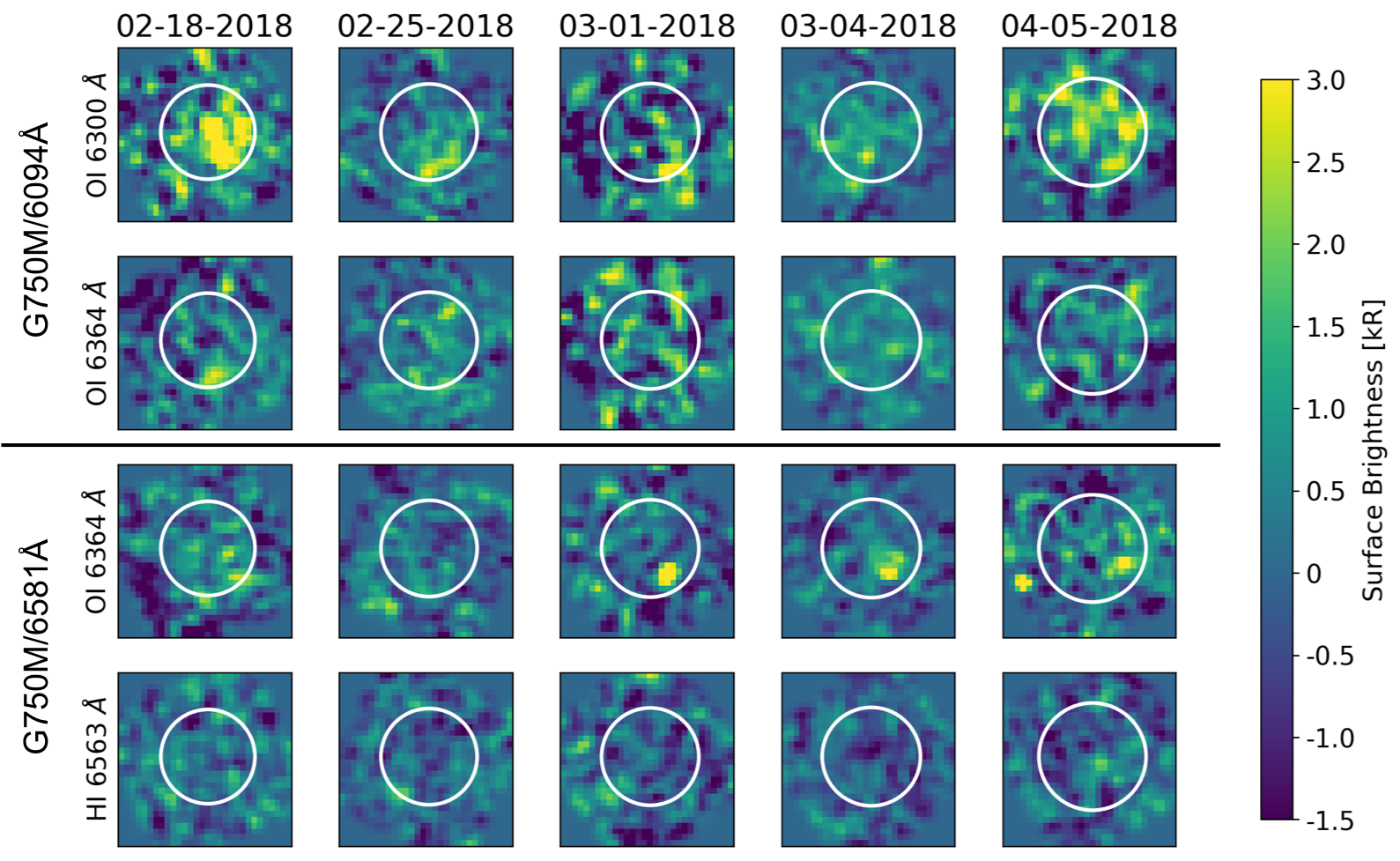}
\caption{HST/STIS images on five dates of the region centered on Europa's location and the OI 6300 and 6364 \AA{} and H$\alpha$ 6563 \AA{} lines. Europa's size and expected location is indicated with a circle, Europa's north pole is up, and all images have been lightly smoothed to bring out extended structures. The upper two rows were taken with the grating centered at 6094 \AA{} and the lower two rows at 6581 \AA{}; OI 6364 \AA{} was covered by both spectral settings. Most small-scale bright features in the images are comparable in intensity to the noise and should be treated with caution. \label{fig:ims}}
\end{figure}

The OI emission image from HST/STIS, averaged across all dates and both wavelengths, is shown in Figure \ref{fig:bigim}. The time-averaged emission is asymmetric and stronger over the trailing/dusk side; the UV aurora have also been observed to be consistently brighter over the dusk side (Roth et al. 2016). Spectral images centered on the OI 6300 and 6364 \AA{} lines are shown in Figure \ref{fig:ims} separately for the five HST observations alongside images centered on the H$\alpha$ 6563 \AA{} line, although H$\alpha$ emission was not detected. Both of the HST/STIS spectral settings covered oxygen emission at 6364 \AA{}, and all individual images at this wavelength are shown to facilitate direct comparison between localized emissions at different wavelengths. Although Europa is spatially resolved, the signal to noise on individual features in the images is low, and such features should be treated with caution. For a given date (total integration time of $\sim$540 seconds), the HST detection limits for emission localized to a 0.2''$\times$0.2'' region are in the range of 0.4-0.8 kR for all dates except March 1 when the limit is 1 kR because of Europa's close proximity to Jupiter on this date. \clearpage
%
\begin{table}
\caption{Disk-Averaged Auroral Brightness\label{tbl:brightness}}
\centering
\begin{tabular}{lllll}
\hline
Date & Tel & OI 6300 \AA{} & OI 6364 \AA{} & H$\alpha$ 6563 \AA{} \\
\hline
2018-02-18 & HST/STIS & 1300$\pm$400 & $<$480 & $<$460 \\
2018-02-25 & HST/STIS &  690$\pm$200 & 490$\pm$150 & $<$200 \\
2018-03-01 & HST/STIS & $<$570 & $<$490 & $<200^a$ \\
2018-03-04 & HST/STIS & 900$\pm$150 & 270$\pm$120 & $<75^a$ \\
2018-03-22 & Keck/HIRES & $>$190 & $>$45 & - \\
2018-04-05 & HST/STIS & 1600$\pm$200 & 800$\pm$150 & $<$80 \\
\hline
\end{tabular}\\
\footnotesize{$^{a}$On 03-01 and 03-04 the derived H$\alpha$ brightness is negative, and 2$\sigma$ upper limits are given. On all other dates 1$\sigma$ upper limits are given.}
\end{table}
\clearpage
%
%
\section{Excitation Model}\label{sec:model}
All of our observations were made in the absence of incident sunlight, and the observed auroral emissions must therefore be produced by electron-impact rather than solar radiation processes. In order to interpret the observed atomic emissions (and non-detections) in terms of the abundances of parent species, we model the excitation of the oxygen O($^1$S) and O($^1$D) and hydrogen Balmer-$\alpha$ states, producing emission at 5577, 6300/6364, and 6563 \AA{} respectively, from electron impact on O, O$_2$, and H$_2$O. For comparison with previous UV observations, we also include excitation of O($^3$S), O($^5$S) and H Lyman $\alpha$ in the model, which emit at 1304, 1356, and 1216 \AA{} respectively. Our model uses the experimentally-derived emission cross-sections described below and plotted in Figure \ref{fig:xsections}, and the electron temperatures and densities derived by Bagenal et al. (2015) from in situ measurements.
\subsection{Cross-sections}\label{sec:xsec}
%
For electron impact dissociation of O$_2$, we use the emission cross-sections measured by Kanik et al. (2003) for production of O($^5$S) and O($^3$S) and those measured by LeClaire and McConkey (1993) for the production of O($^1$S). The cross-sections for O($^1$D) have not been measured, but Suzuki et al. (2011) give cross-sections for electron-impact excitation of the Schumann-Runge continuum. The dominant decay path for such excited O$_2$ molecules is through O($^3$P)$+$O($^1$D), and for our model we therefore use these cross-sections, assuming that 90\% of the excited O$_2$ molecules produce an atom in the O($^1$D) state (Suzuki et al. 2011; Cosby et al. 1993), and add this to the cascade contribution from O($^1$S).\par
%
%
For electron impact excitation of atomic O, we use the emission cross-sections measured by Zipf and Erdman (1985) for production of O($^3$S). For production of O($^5$S), Doering and Gulcicek (1989b) provide excitation cross-sections, and estimate that the total emission cross section peaks a factor of three higher due to cascades from higher-energy states, dominated by O($^5$P). We therefore use their excitation cross sections increased by a factor of three, though we note that the effect of the cascade contributions to the shape of the cross-section energy dependence is not accounted for. For production of O($^1$S) we use the emission cross-sections measured by Laher and Gilmore (1990) and the excitation cross-sections for O($^1$D) measured by Doering and Gulcicek (1989a) and updated by Doering and Gulcicek (1992), to which we add the cascade contribution from O($^1$S). \par
%
For electron-impact dissociative excitation of H$_2$O, we use the emission cross-sections measured by Kedzierski et al. (1998) for O($^1$S), and the cross-sections measured by Morgan and Mentall (1974) renormalized by Itikawa and Mason (2005) for O($^3$S). The cross-section for O($^5$S) is estimated to be 25\% that of O($^3$S) at 100 eV (Makarov et al. 2004), and we use this scaling to calculate cross sections for O($^5$S) emission. There are no experimental cross-sections in the literature for production of O($^1$D) from the electron-impact dissociation of H$_2$O. Bhardwaj and Raghuram (2012) use analytical calculations by Jackman et al. (1977) to estimate a branching ratio for oxygen produced by this process, and we adopt their approach here to compute the O($^1$D) rates.\par
For the atomic hydrogen lines, Itikawa and Mason (2005) compile experimental cross-sections for hydrogen emission due to electron impact dissociation of H$_2$O, renormalizing older work to more recent reference values; our choices follow their recommendations. In the UV, we use the emission cross-sections presented by Makarov et al. (2004) for Ly$\alpha$ emission at 1216 \AA{}. Beenakker et al. (1974) measured the four lowest-energy Balmer emissions and presented emission cross-sections for H$\beta$ along with scale factors to scale to other transitions, noting that the energy dependence is similar for all emissions. We use their results for H$\alpha$ emission (6563 \AA{}). \par
\begin{figure}[ht]
\centering
\includegraphics[width=14cm]{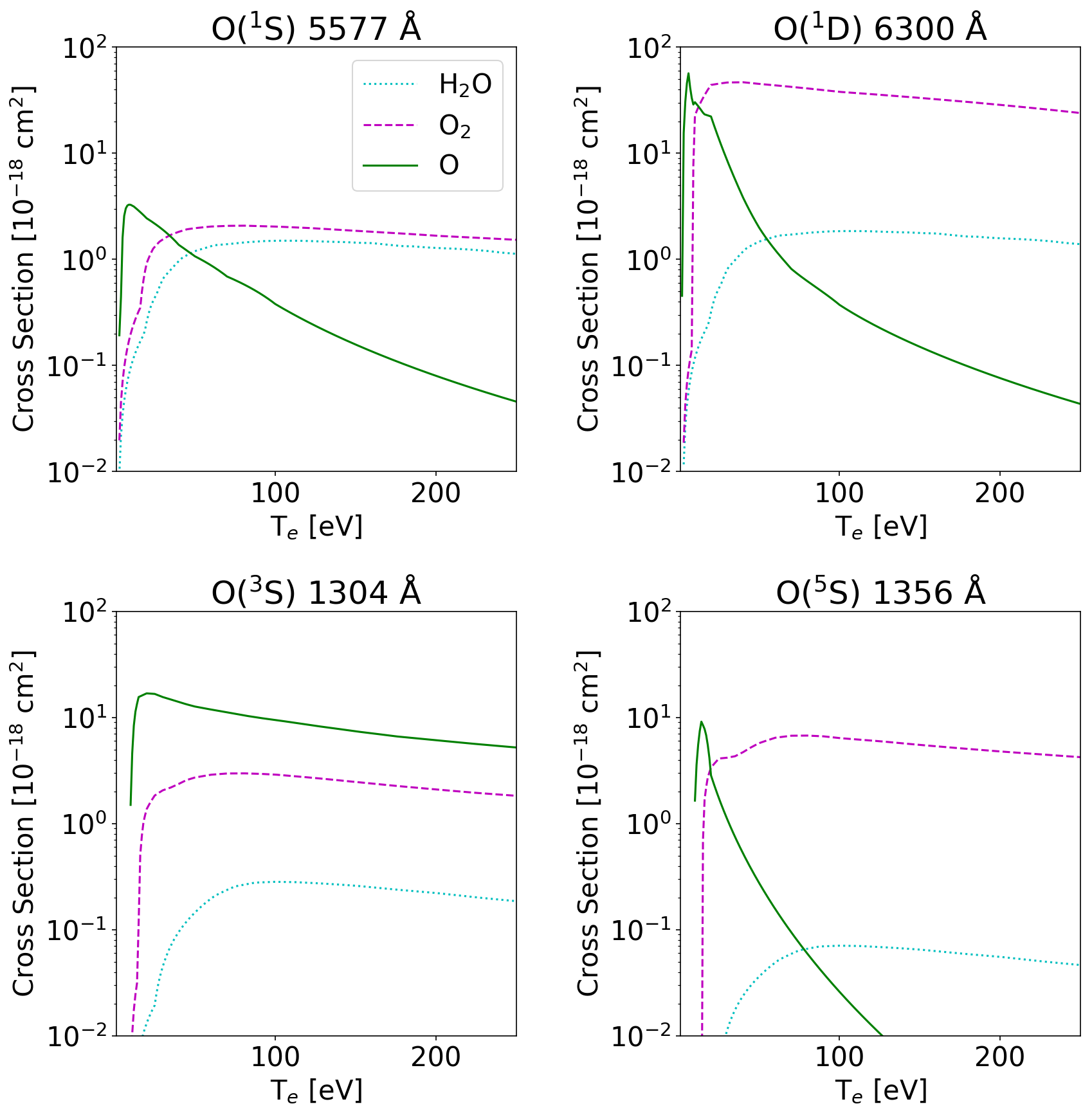}
\caption{Emission cross-sections for UV and visible oxygen emission lines following electron impact on O, O$_2$, and H$_2$O, as described in Section \ref{sec:xsec}. The corresponding emission rates calculated for the plasma conditions at Europa's orbit are given in Table \ref{tbl:rates}. \label{fig:xsections}}
\end{figure}

When the available cross-section data do not extend to high enough electron energies, we assume an exponential drop-off in cross section with electron energy. All oxygen cross-sections used in our model are shown in Figure \ref{fig:xsections}. \par
\subsection{Emission rates}
We use the experimental cross-sections described above to determine emission rates from excited states of OI and HI for plasma conditions at Europa's orbit. The rate $\gamma$ is found by integrating the cross section over all electron energies able to excite the state, for a given electron energy distribution: \\
\begin{equation}
\gamma = \int_\chi^\infty v\sigma (v) f(v) dv
\end{equation}
where $\chi$ is the threshold electron energy, $\sigma (v)$ is the energy-dependent cross-section, and $f(v)$ is the electron velocity distribution. For our calculations we adopt two Maxwellian velocity distributions centered at 20 and 250 eV, with the latter constituting 5\% of the total population, appropriate for the plasma conditions at Europa's orbit (Bagenal et al. 2015) and consistent with previous work (Roth et al. 2016). Our calculated rates are consistent with those derived for the UV transitions by Roth et al. (2016) using similar methods. The final photon rate from the $i$th species, in units of photons/s/cm$^{3}$ can then be calculated as $n_e n_i \gamma_i$, where $n_e$ and $n_i$ are the particle densities of electrons and species $i$ in units of cm$^{-3}$.\par
The calculated rates for each transition, integrated across the electron energy distribution, are tabulated in Table \ref{tbl:rates}. We use these rates to estimate the column density of the gas in the case of both an O and an O$_2$ atmosphere using the median value for the electron density $n_e$ at Europa's orbit of 160 cm$^{-3}$ (Bagenal et al. 2015). However, as the electron density varies by a factor of almost 20, the derived column densities are highly uncertain and should treated as order-of-magnitude estimates. \par
%
\begin{table}
\caption{Emission rate coefficients for plasma conditions at Europa's orbit \label{tbl:rates}}
\centering
\begin{tabular}{lllll}
\hline
Species & Wavelength & O & O$_2$ & H$_2$O \\
 & [\AA{}] & [cm$^3$ s$^{-1}$] & [cm$^3$ s$^{-1}$] & [cm$^3$ s$^{-1}$] \\
\hline
O($^1$S) & 5577 & 4.8$\times$10$^{-10}$ & 4.4$\times$10$^{-10}$ & 2.5$\times$10$^{-10}$ \\
O($^1$D) & 6300/6364 & 3.3$\times$10$^{-9}$ & 1.2$\times$10$^{-8}$ & 3.1$\times$10$^{-10}$  \\
O($^3$S) & 1304 & 3.7$\times$10$^{-9}$ & 5.9$\times$10$^{-10}$ & 3.3$\times$10$^{-11}$ \\
O($^5$S) & 1356 & 4.6$\times$10$^{-10}$ & 1.3$\times$10$^{-9}$ & 8.2$\times$10$^{-12}$ \\
Ly$\alpha$ & 1216 & - & - & 9.3$\times$10$^{-10}$ \\
H$\alpha$ & 6563 & - & - & 5.0$\times$10$^{-10}$ \\
\hline
\end{tabular}
\end{table}
\section{Discussion} \label{sec:disc}
\subsection{Mixing ratios and parent molecules} \label{sec:uvoptratio}
The ratios of auroral emission lines tracing different atomic excitation states have been used for decades to infer the presence and mixing ratios of parent atoms and molecules in the atmospheres of Europa and other satellites. In the UV, the ratio of 1356/1304 \AA{} oxygen emission has been repeatedly measured in the range of 1-3, indicative of a predominantly O$_2$ atmosphere with at most a few percent atomic O (Hall et al. 1998; Roth et al. 2016). A localized variation in the UV oxygen emission ratio was also one of the lines of evidence indicative of an H$_2$O plume in 2012 UV data (Roth et al. 2014). Atomic oxygen emits at several wavelengths in the 5000-9000 \AA{} range, and the ratios of these species to one another and to the UV lines provide an independent assessment of parent molecules and atmospheric mixing ratios. Table \ref{tbl:ratios} shows the expected intensity ratios for hydrogen and oxygen auroral emission lines in the UV and visible, calculated from the model described in Section \ref{sec:model} for parent species O, O$_2$, and H$_2$O. \par
%
\begin{table}
\caption{Emission Ratios given plasma conditions at Europa's orbit\label{tbl:ratios}}
\centering
\begin{tabular}{llllll}
\hline
Species & Wavelengths [$\mathrm{\AA}$] & O & O$_2$ & H$_2$O \\
\hline
Ly$\alpha$/H$\alpha$ & 1216/6563 & - & - & 1.9 \\
O($^1$D)/O($^1$S) & (6300+6364)/5577 & 6.7 & 27 & 1.2 \\
O($^1$D)/O($^5$S) & (6300+6364)/1356 & 7.1 & 9.2 & 37 \\
O($^1$D)/O($^3$S) & (6300+6364)/1304 & 0.9 & 20 & 9.3 \\
O($^5$S)/O($^3$S) & 1356/1304 & 0.1 & 2.2 & 0.25 \\
\end{tabular}\\
\end{table}
We detected the oxygen red doublet at 6300/6364 \AA{}, which arises from the O($^1$D)$\rightarrow$O($^3$P) transition, at an average disk-integrated brightness of 1.2 kR. Using the Keck HIRES dataset, including the non-detection of 5577 \AA{} emission arising from O($^1$S), we place a lower limit of 21 on the O($^1$D)/O($^1$S) ratio (i.e. ratio of emission at 6300$+$6364 \AA{} to 5577 \AA{}). The expected ratio is 27 for a purely O$_2$ atmosphere and 6.7 for a purely O \textbf{atmosphere} (see Table \ref{tbl:ratios}). The observations therefore favor an O$_2$ atmosphere, with an upper limit on the O/O$_2$ mixing ratio of 0.35. The lower limit on the O($^1$D)/O($^1$S) ratio is based on the best few integrations from the Keck dataset; if this experiment were repeated during clear weather conditions and all spectra were usable, the noise level would be a factor of $\sim$2-3 lower and we would expect to see 5577 \AA{} emission for either parent molecule, which would provide a stronger constraint on the O/O$_2$ mixing ratio. \par
The O($^1$D)/O($^5$S) emission ratio is also weakly diagnostic of the O/O$_2$ mixing ratio; ratios of seven and nine are expected for atomic and molecular oxygen respectively. Roth et al. (2016) measured an average disk-integrated O($^5$S) emission of 80 R, while we measure an average disk-integrated O($^1$D) brightness of 1.2 kR, a factor of 15 higher. However, we note that the O($^1$D) emission varies over $\sim$0.5-2 kR, while Roth et al. (2016) saw variability in the UV aurorae from $\sim$35 to $>$150 R. Thus while the ratio of average visible to UV emission from the measurements is higher than predicted for either an O or O$_2$ atmosphere, a meaningful comparison would require measuring the UV and visible aurorae simultaneously. \par
\begin{figure}[ht]
\centering
\includegraphics[width=14cm]{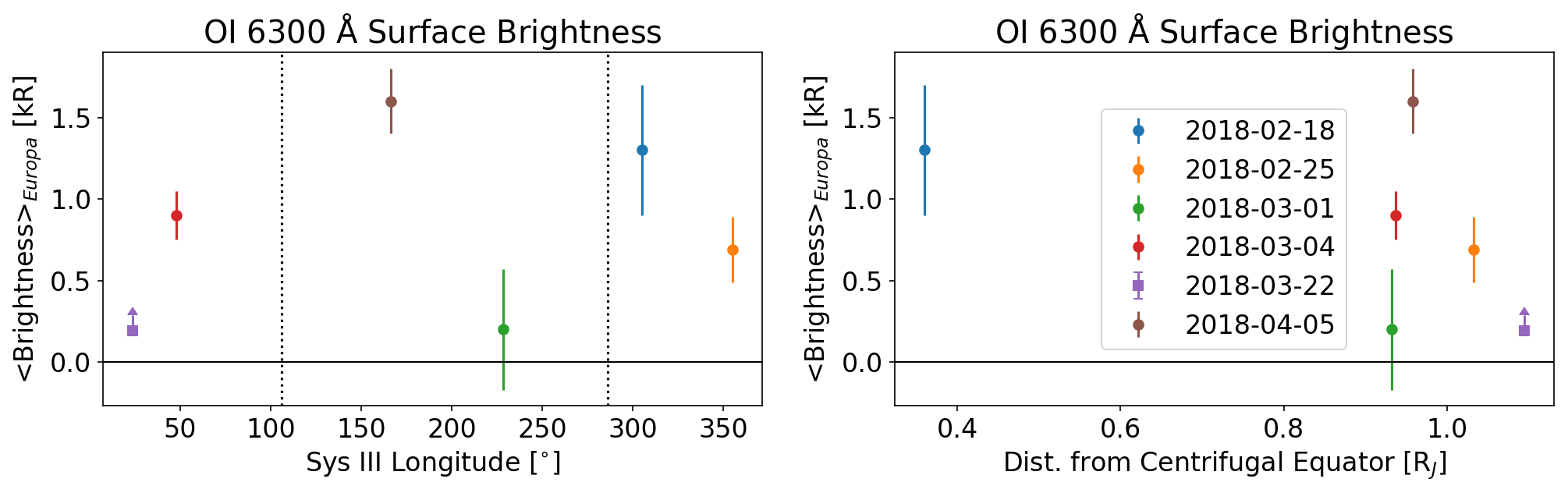}
\caption{Disk-averaged surface brightness of Europa's 6300 \AA{} emission on six dates based on HST/STIS (circles) and Keck/HIRES (square) observations, as a function of Europa's position in Jupiter's System III longitude, and of distance from Jupiter's centrifugal equator. The auroral activity is expected to be higher at smaller centrifugal equator distances, and at System III longitudes closer to Europa's plasma sheet crossings, indicated on the left plot as dotted vertical lines. \label{fig:var}}
\end{figure}

Roth et al. (2016) attribute much of the observed variability in the UV dataset to Europa's position relative to the plasma sheet, finding only a $\sim$20\% variation from their average value after correcting for this effect. Figure \ref{fig:var} shows the 6300 \AA{} brightness from our dataset as a function of Europa's position in Jupiter's System III longitude at the time of the observations, and as a function of distance from the centrifugal equator. We find no significant correlation with either of these parameters, but note that five of our six observations were made when Europa was nearly at maximal distance from the plasma sheet. \par
We convert the measured surface brightnesses to column densities using an electron density of 160 cm$^{-3}$, an average value for the plasma at Europa's orbit (Bagenal et al. 2015). The electron density varies temporally and spatially by more than an order of magnitude, and our derived column densities are thus similarly uncertain. We find column densities of 5-30$\times$10$^{14}$ cm$^{-2}$ for a pure atomic oxygen atmosphere, and 1-9$\times$10$^{14}$ cm$^{-2}$ for a pure molecular oxygen atmosphere. While these column densities match well with early estimates for an O$_2$ atmosphere based on the UV aurorae (Hall et al. 1998), those models assumed an electron density of 30-50 cm$^{-3}$. For an electron density of 40 cm$^{-3}$, our measured surface brightnesses would correspond to an O$_2$ column density of 4-35$\times$10$^{14}$ cm$^{-2}$, which is consistent with past UV results for the lower end of the range but exceeds them by a factor of $\sim$2 at the upper end of the range. \par
\subsection{Potential for detecting water plume activity from the optical aurora} \label{sec:hydrogen}
In December 2012, Roth et al. (2014) detected an enhancement in Lyman-$\alpha$ emission, co-located with an enhancement in OI 1304 \AA{} but not OI 1356 \AA{} emission, off the limb of Europa. These features are together characteristic of dissociative excitation of H$_2$O (see Table \ref{tbl:ratios}), and the authors attributed the emission to a water plume. Here we discuss the potential for detection of water plumes from the visible-wavelength emissions using ground- and space-based telescopes. \par
The visible-wavelength oxygen emissions at 6300 and 5577 \AA{} are a clearer diagnostic for identifying localized H$_2$O than the UV emissions at 1304/1356 \AA{}. While the UV line ratio produced by dissociation of H$_2$O differs from that of O$_2$ by a factor of 10, it falls between the expected ratios for O and O$_2$. The presence of H$_2$O would therefore have a similar effect as a higher O/O$_2$ mixing ratio, which introduces some ambiguity in interpretation, particularly in the off-limb regions that fall within the extended corona, where the mixing ratio may be as high as 35\% compared to $<$5\% near the surface (Roth et al. 2016). In contrast, for the visible-wavelength emissions any combination of O and O$_2$ will produce a (6300+6364)/5577 \AA{} line strength ratio between 6 and 30, and the ratio of 1.2 produced by H$_2$O dissociation would therefore provide an unambiguous indicator, particularly if localized to a region where H$\alpha$ emission and an enhancement in 6300 \AA{} were also detected.\par
For the localized enhancements in the 1304 \AA{} and Ly$\alpha$ surface brightness attributed to an H$_2$O plume by Roth et al. (2014), the corresponding surface brightness for the visible emissions from OI at 6300 \& 5577 \AA{} and H$\alpha$ at 6563 \AA{} would be 200-250 R. Based on the noise level in our HST dataset over spatial regions of the same size, we would detect such emission at the 2$\sigma$ level after 100-150 minutes of integration time. While this is comparable to the amount of observing time required to detect the corresponding emissions at UV wavelengths, optical observations must be made while Europa is in eclipse and each visit is therefore limited to the $\lesssim$2 hour eclipse duration. The potential for detecting water plume activity in HST observations of the visible aurora therefore depends strongly on whether the plumes remain active at a detectable level for the several days over which such observations would need to be made.\par
Detection of water plumes from a ground-based telescope such as Keck requires a signal that is diagnostic of H$_2$O even in a disk-integrated spectrum, since the typical seeing is at the same angular scale as Europa (0.8-1.0''). The disk-averaged H$\alpha$ brightness of 15 R that corresponds to the localized emission level discussed above would be detected at the 2$\sigma$ level after 35 minutes of on-source time based on the noise level in our single best 5-minute integration with Keck/HIRES. The corresponding enhancement in O($^1$S) 5577 \AA{} emission would also be detectable through the change in the 6300/5577 \AA{} ratio: an H$_2$O plume with localized column density of 1.5$\times10^{16}$ cm$^{-2}$ (Roth et al. 2014) would lower the disk-integrated line ratio from 21 to 13, assuming an O$_2$ column density of 5$\times$10$^{14}$ cm$^{-2}$. While a change in the O/O$_2$ mixing ratio would also lower this emission ratio, increasing the mixing ratio from $<$1\% up to 5\%, the highest O percentage inferred in the near-surface atmosphere by Roth et al. (2016), would only decrease the emission line ratio from 21 to 20. The presence of a water plume would therefore be distinguishable from a reasonable change to the O/O$_2$ mixing ratio at the 2$\sigma$ level in Keck spectra after 60-80 minutes of integration time. A detection of H$\alpha$ emission, combined with an enhancement in O($^1$S) emission relative to O($^1$D), would therefore provide strong evidence for an H$_2$O plume; such a detection would be possible during a single eclipse observed with Keck/HIRES, for the plume properties inferred by Roth et al. (2014). \par
\section{Conclusions} \label{sec:conc}
%
We present the first detection of Europa's 6300 \& 6364 \AA{} oxygen aurora, which is produced predominantly by electron-impact dissociation of O$_2$ in the atmosphere. The detections are based on optical spectra of Europa in eclipse by Jupiter obtained on six occasions in the spring of 2018 with the Keck and Hubble Space Telescopes. No other auroral emission lines are detected, including sodium and hydrogen H$\alpha$. The surface brightness of the O($^1$D) 6300 \AA{} line, in combination with previous observations of UV oxygen emissions (Roth et al. 2016) and the upper limit on O($^1$S) based on the non-detection of the 5577 \AA{} line, favors an O$_2$ atmosphere with a column density of 10$^{14}$-10$^{15}$ cm$^{-2}$, consistent with results from past UV observations (Hall et al. 1998; Roth et al. 2016). \par
We present expected ratios for the UV and visible oxygen and hydrogen lines given a particular parent atom or molecule, adding to the ratios used in the UV. The 6300/5577 \AA{} ratio in particular can provide a strong independent diagnostic of the O/O$_2$ mixing ratio to support the UV results. The 5577 \AA{} emission was not detected in our dataset, and the upper limit on O($^1$S)/O($^1$D) of 21 rules out an O/O$_2$ mixing ratio above 0.35; a future similar Keck campaign during clear conditions should detect the 5577 \AA{} line and provide a stronger constraint on the atmospheric composition. \par
In the visible regime, the presence of H$\alpha$ emission at 6563 \AA{} and the ratio of oxygen 6300/5577 \AA{} emission serve as a diagnostic of the presence of water. Using the excitation model and noise measurements derived from the datasets presented here, we estimate that the atomic H and O signatures produced by water plume activity would be detectable in Keck observations of a single Europa eclipse, while a localized 2$\sigma$ detection with HST/STIS would require observations spanning multiple eclipses and therefore only be sensitive to a water plume that remains active at a detectable level for at least 5-10 days. \par
%
\clearpage
\section*{Acknowledgements}
The authors are grateful to Samantha Trumbo for assistance in obtaining the Keck data. K. de Kleer is supported by a Heising-Simons Foundation \textit{51 Pegasi b} postdoctoral fellowship. Support for this work was also provided by NASA through grant number HST-GO-15425.002-A from the Space Telescope Science Institute, which is operated by AURA, Inc., under NASA contract NAS 5-26555. This work was based in part on observations made with the NASA/ESA Hubble Space Telescope, obtained from the data archive at the Space Telescope Science Institute. Some of the data presented herein were obtained at the W. M. Keck Observatory, which is operated as a scientific partnership among the California Institute of Technology, the University of California and the National Aeronautics and Space Administration. The Observatory was made possible by the generous financial support of the W. M. Keck Foundation. The authors wish to recognize and acknowledge the very significant cultural role and reverence that the summit of Maunakea has always had within the indigenous Hawaiian community.  We are most fortunate to have the opportunity to conduct observations from this mountain. This work made use of the JPL Solar System Dynamics high-precision ephemerides through the HORIZONS system.
\clearpage
\section*{References}
\begin{itemize}
\item[]{Bagenal, F. et al. 2015. Plasma conditions at Europa's orbit. Icarus 261, 1-13.}
\item[]{Beenakker, C.I.M., de Heer, F.J., Krop, H.B., M{\"o}hlmann, G.R. Chem. Phys. 6, 445 (1974).}
\item[]{Bhardwaj, A. and Raghuram, S. et al. 2012. A coupled chemistry-emission model for atomic oxygen green and red-doublet emissions in the comet C/1996 B2 Hyakutake. ApJ 748:13 (18pp).}
\item[]{Brown, M.E. and Hill, R.E., 1996. Discovery of an extended sodium atmosphere around Europa. Nature 380, 229-231.}
\item[]{Brown, M.E. 2001. Potassium in Europa's atmosphere. Icarus 11, 190-195.}
\item[]{Cosby, P.C. 1993. Electron-impact dissociation of oxygen. J. Chem. Phys. 98, 7804 9544; 9560.}
\item[]{Doering, J.P. and Gulcicek, E.E., 1989a. Absolute differential and integral electron excitation cross sections for atomic oxygen VII. The $^3P\rightarrow^1D$ and $^3P\rightarrow^1S$ transitions from 4.0 to 30 eV. JGR 94, 1541-1546.}
\item[]{Doering, J.P. and Gulcicek, E.E., 1989b. Absolute differential and integral electron excitation cross sections for atomic oxygen VIII. The $^3P\rightarrow^5S^0$ transition (1356 \AA{}) from 13.9 to 30 eV. JGR 94, 2733-2736.}
\item[]{Doering, J.P. 1992. Absolute differential and integral electron excitation cross sections for atomic oxygen 9. Improved cross section for the $^3$P$\rightarrow ^1$D transition from 4.0 to 30 eV. JGR 97, 19531-19534.}
\item[]{Hall, D.T., Feldman, P.D., McGrath, M.A., Strobel, D.F., 1998. The far-ultraviolet oxygen aurora of Europa and Ganymede. ApJ 499, 475-481.}
\item[]{Hall, D.T., Strobel, D.F., Feldman, P.D., McGrath, M.A., Weaver, H.A., 1995. Detection of an oxygen atmosphere on Jupiter's moon Europa. Nature 373, 677-679.}
\item[]{Itikawa, Y. and Mason, N. 2005. Cross sections for electron collisions with water molecules. J. Phys. Chem. Ref. Data 34, 1-22.}
\item[]{Jackman, C.H., Garvey, R.H., Green, A.E.S. 1977. J. Geophys. Res. 82, 5081.}
\item[]{Jia. X., Kivelson, M.G., Khurana, K.K., Kurth, W.S., 2018. Evidence of a plume on Europa from Galileo magnetic and plasma wave signatures. Nature Astronomy Letters. doi:10.1038/s41550-018-0450-z}
\item[]{Kanik, I. et al. 2003. Electron impact dissociative excitation of O$_2$: 2. Absolute emission cross sections of the OI(130.4 nm) and OI(135.6 nm) lines. JGR 108, E11, CiteID 5126.}
\item[]{Kedzierski, W., J. Derbyshire, C. Malone, and J. W. McConkey. Isotope effects in the electron impact break-up of water. J. Phys. B 31, 5361 (1998).}
\item[]{Laher, R.R. and Gilmore, F.R. 1990. Updated excitation and ionization cross sections for electron impact on atomic oxygen. J. Phys. Chem. Ref. Data 19, 277-305.}
\item[]{LeClair, L.R. and McConkey, J.W. 1993. Selective detection of O($^1S_0$) following electron impact dissociation of O$_2$ and N$_2$O using a XeO$^*$ conversion technique. Journal of Chem. Phys. 99, 4566-4577.}
\item[]{Makarov, O., et al. 2004. Kinetic energy distributions and line profile measurements of dissociation products of water upon electron impact. JGR 109, A09303, p. 1-15.}
\item[]{McCord, T.B. et al. 1999. Hydrated salt minerals on Europa's surface from the \textit{Galileo} Near Infrared Mapping Spectrometer (NIMS) investigation. JGR 104, 11827-11852.}
\item[]{McGrath, M.A., Sparks, W.B., 2017. Galileo ionosphere profile coincident with repeat plume detection at Europa. Research Notes at the AAS, 1, 1.}
\item[]{Morgan, H.D. and J. E. Mentall, J. Chem. Phys. 60, 4734 (1974).}
\item[]{Roth, L., et al. 2014. Science 343, 171-174.}
\item[]{Roth, L., et al. 2016. JGR: Space Physics 121, 2143-2170.}
\item[]{Roth, L., et al. 2017. AJ 153:67 (10pp).}
\item[]{Sparks, W.B. et al. 2017. Active cryovolcanism on Europa? ApJ 839, L18, p. 1-5.}
\item[]{Sparks, W.B. et al. 2016. Probing for evidence of plumes on Europa with HST/STIS. ApJ 829, 121, p. 1-21.}
\item[]{Suzuki, D. et al. 2011. J Chem Phys 134, 064311, 1-8.}
\item[]{Vogt, S.S. et al., 1994, Proc. SPIE, 2198, 362}
\item[]{Woodman, J.H., Cochran, W.D., Slavsky, D.B., 1979. Spatially resolved reflectivities of Jupiter during the 1976 opposition. Icarus 37, 73-83.}
\item[]{Zipf, E.C. and P.W. Erdman, 1985. J Geophys. Res. 90, 11087.}
\end{itemize}



\end{document}